\documentclass[a4paper,fleqn]{cas-dc}
\usepackage{graphicx,subcaption}
\usepackage{multicol}
\usepackage{layout}
\usepackage{amsfonts}
\usepackage{amsmath}
\usepackage{amssymb}
\usepackage{textcomp}
\usepackage{caption}
\usepackage[normalem]{ulem}
\usepackage[utf8]{inputenc}
\usepackage{natbib} 
\def\tsc#1{\csdef{#1}{\textsc{\lowercase{#1}}\xspace}}
\tsc{WGM}
\tsc{QE}

\newcommand{\lum}{erg\,s$^{-1}$}

\newcommand{\phflux}{\mbox{${\rm \, ph \,\, cm^{-2} \, s^{-1}}$}}

\newcommand{\gm}{$\gamma$}

\begin{document}
\UseRawInputEncoding
\let\WriteBookmarks\relax
\def\floatpagepagefraction{1}
\def\textpagefraction{.001}

\shorttitle{VHE emission from misaligned jets}
\shortauthors{Tomar et al.}

\title[mode = title]{\textit{Fermi}-Large Area Telescope Detection of Very High Energy ($>$100 GeV) Emission from Misaligned Jetted Active Galactic Nuclei}

\author[1]{Gunjan Tomar}[orcid=0000-0003-4992-6827]
\affiliation[1]{organization={Astronomy \& Astrophysics group, Raman Research Institute},
            addressline={C. V. Raman Avenue, 5th Cross Road, Sadashivanagar}, 
            city={Bengaluru},
            postcode={560080}, 
            state={Karnataka},
            country={India}} 

\author[2]{Vaidehi S. Paliya}[orcid=0000-0001-7774-5308]
\ead{vaidehi.s.paliya@gmail.com}
\affiliation[2]{organization={Inter-University Centre for Astronomy and Astrophysics (IUCAA)},
            addressline={SPPU Campus}, 
            city={Pune},
            postcode={411007}, 
            state={Maharashtra},
            country={India}} 
\cormark[1]

\author[2]{D. J. Saikia}[orcid=0000-0002-4464-8023]

\author[3]{C. S. Stalin}[orcid=0000-0002-4998-1861]
\affiliation[3]{organization={Indian Institute of Astrophysics},
            addressline={Block II, Koramangala}, 
            city={Bengaluru},
            postcode={560034}, 
            state={Karnataka},
            country={India}} 

\cortext[cor1]{Corresponding author}
\nonumnote{}
\begin{abstract}
The detection of very-high-energy (VHE; $>$100 GeV) $\gamma$-ray radiation from misaligned jetted Active Galactic Nuclei (AGN) challenges the emission models that primarily explain VHE emissions from beamed AGN, i.e., blazars. Using over 16 years of \textit{Fermi}-Large Area Telescope (\textit{Fermi}-LAT) Pass 8 data in the energy range 0.1$-$2 TeV, we systematically explore the VHE emission from a recently published sample of 160 radio galaxies. We identify 12 sources detected at $>4\sigma$ confidence level (test statistic or TS$>$16), including nine with TS$>$25 and two Fanaroff-Riley type II objects. This detected sample includes seven out of eight previously known VHE objects. Two radio galaxies are detected in the VHE band for the first time, and we identify three promising candidates with 16$<$TS$<$25. Additionally, 13 objects are identified as candidate VHE emitters with at least one VHE photon detected with the \textit{Fermi}-LAT. These findings expand the sample of known VHE-emitting radio galaxies, whose multiwavelength follow-up observations can help provide insights into the emission mechanisms, jet physics, and the contribution of misaligned AGN to the extragalactic $\gamma$-ray background.
\end{abstract}


\maketitle

\section{Introduction}

The \textit{Fermi} Large Area Telescope (LAT) has revolutionized our knowledge of the \gm-ray sky at high energies (HE; 100 MeV $-$ 100 GeV), especially in studying the emissions from Active Galactic Nuclei (AGN). These objects are powered by accretion onto supermassive black holes (SMBHs) at their centers, emit across the entire electromagnetic spectrum, and often host powerful, relativistic jets \citep[cf.][]{1995PASP..107..803U}. Blazars, with the jets aligned to our line of sight, dominate the sky at HE and very high energies (VHE; $>$100 GeV), despite being only $\lesssim1\%$ of the jetted AGN population. On the other hand, radio galaxies, considered the parent population of blazars, are the most numerous jetted AGN, although only a few are detected in the HE and VHE bands. Owing to the misalignment of their jets with our line of sight, their detection at GeV$-$TeV energies is surprising since they have either no or weak Doppler boosting that enhances the flux in blazars and allows their detection. 
\par
\textit{Fermi}-LAT has been surveying the entire sky in the 20 MeV to over 2 TeV energy range since August 2008 and, therefore, provides the most extensive catalogs of \gm-ray sources. The latest update of the fourth source catalog (4FGL-DR4) covers over 14 years and contains a total of 7,195 sources in the energy range of 50 MeV to 1 TeV \citep[][]{2020ApJS..247...33A,4fgl_dr4}. About $\sim$56$\%$ of these sources are associated with AGN, including blazars, Narrow-line-Seyfert 1 galaxies, and misaligned AGN. Blazars dominate this sample with 3,933 sources associated with flat spectrum radio quasars, BL Lacs, and blazars of unknown type (BCU). With only 53 sources, radio galaxies form only a tiny fraction of this population of \gm-ray emitting AGN. 
\par
At VHE energies, there are 91 sources associated with AGN that have been detected by the ground-based Imaging Atmospheric Cherenkov telescopes (IACTs), including the High Energy Stereoscopic System (H.E.S.S.), the Major Atmospheric Gamma Imaging Cherenkov Telescopes (MAGIC), and the Very Energetic Radiation Imaging Telescope Array System (VERITAS) as listed in TeVCat\footnote{As on 28 October 2024.} \citep{tevcat08}. Only six of these are known to be associated with radio galaxies, namely NGC~1275 ($z=0.018$), 3C~264 ($z=0.021$), M87 ($z=0.004$), Centaurus~A ($z=0.002$), IC~310 ($z=0.019$), and PKS 0625$-$35 ($z=0.055$). \citet[][]{angioni_20} performed simulations of observations with the upcoming Cherenkov Telescope Array Observatory (CTAO) and predicted VHE detection of 11 radio galaxies \citep[see also,][]{2020ApJ...891..145B}. The detection of radio galaxies at these energies, despite their large viewing angle, is intriguing and could hint at different emission sites and processes than in blazars. For instance, HE and VHE emission from the extended scales of the nearest radio galaxy, Centaurus A, have been localized by \textit{Fermi}-LAT \citep{fermi_cena_ext_2010} and H.E.S.S. \citep{cena_hess_2020}, respectively. Recently, \citet[][]{2024ApJ...965..163Y} reported the detection of \gm-ray emission from the lobes of a giant radio galaxy, NGC 6251 ($z=0.02$). These findings have major implications for our understanding of the \gm-ray production mechanism and the location of the \gm-ray emitting region in the sources \citep[see, e.g.,][for a recent review]{2022Galax..10...61R}. 
\par
Several models have been proposed to explain the HE and VHE emissions from radio galaxies, indicating the complexities of their emission locations and processes \citep[see,][for a review]{reiger2018}. For example, a homogeneous single-zone Synchrotron Self-Compton (SSC) model is widely used to explain the HE-VHE emission in blazars \citep[e.g.,][]{2008ApJ...686..181F}. The same model has been successful in explaining the emission from the radio galaxy NGC~1275 during flaring and low-activity states with a modest Doppler factor of $\sim$3 \citep[see, e.g.,][]{godambe24_ngc1275}. However, this model is often found to be inadequate for other radio galaxies. For example, it fails to fully reproduce the spectral energy distribution (SED) of Centaurus~A and 3C~264, both of which require an additional emission component \citep{sahakyan13, 3c264_sed_19}. Additionally, the rising TeV component in IC~310 suggests the need for multiple electron populations or a hadronic origin \citep{fraija17}. The fast $\sim$minute-scale TeV variability in IC~310 has been attributed to $\gamma$-ray emission from a compact region close to the SMBH, aligned along the line of sight before the jet bends to a larger viewing angle \citep{2020MNRAS.499.5791G}. Additionally, $\gamma$-ray emission from the spine-sheath-structured jets has also been proposed for radio galaxies like NGC~1275 and 3C~264, which is supported by the presence of limb-brightened jets in these sources \citep[cf.][]{georganopoulos_spinesheath_2003,ghisellini_spinesheath_2005}. To explain the rapid variability, as in the case of NGC~1275 and M87, models such as emission from magnetospheric gaps \citep{ngc1275_magic18}, mini-jet, a photo-hadronic model \citep{alfaro22}, or an external cloud(/star)-jet interaction model \citep{cloud_jet2} have also been proposed. 
\par
These examples demonstrate the complexities involved in explaining the origin of HE and VHE emission from misaligned jetted AGN. Since the viewing angle of the jet is large, these objects enable us to probe the radiative mechanisms responsible for the observed emission from a different vantage point than blazars. However, given the small sample of known HE and VHE-emitting radio galaxies, it is difficult to carry out population studies and draw meaningful constraints on the physics of relativistic jets. Clearly, the need of the hour is to increase the sample size of HE and VHE emitting misaligned jetted AGN. Particularly, identifying VHE-emitting radio galaxies is crucial, since they can be used to study some of the most efficient particle accelerators \citep[see, e.g.,][]{2020NatAs...4..124B}. Increasing the sample size of the VHE-detected misaligned AGN is also critical to constrain their contribution to the extragalactic \gm-ray background \citep[e.g.,][]{2014ApJ...780..161D,2019ApJ...879...68S,2022ApJ...931..138F}. An enlarged sample of VHE-emitting radio galaxies will also permit us to probe the surrounding environment in which jets are launched and grow \citep[cf.][]{2020MNRAS.499.5791G,2025MNRAS.536.2025M}. Furthermore, increasing the sample size of these objects will also allow us to identify the best targets for the upcoming CTAO \citep[e.g.,][]{2022MNRAS.512..137N}.
\par
Ground-based IACTs such as H.E.S.S., MAGIC, and VERITAS are designed to detect $\gamma$-rays extending up to energies of tens of TeV, with high sensitivity and angular resolution. However, they have limitations due to restricted sky coverage, dependence on weather conditions, clear night skies, and source visibility, as observations are limited to specific times and locations. On the other hand, \textit{Fermi}-LAT's ability to continuously survey the entire sky enables near-uninterrupted monitoring that is unaffected by weather or ground-based visibility constraints. It operates in an energy range of 20 MeV to over 2 TeV. Although its sensitivity declines at higher energies, the reduced background contamination at these levels improves source localization and photon association. This allows for a complementary approach to identify and study VHE-emitting radio galaxies. In this work, we use the \textit{Fermi}-LAT data above 100 GeV to identify the brightest VHE-emitting radio galaxies from the latest \textit{Fermi} catalog. The details about the sample we use in this study are discussed in Section \ref{sec:sample}. The methodology for identification is highlighted in Section \ref{sec:fermi}. The results and the notes on individual objects are presented in Sections \ref{sec:results} and \ref{sec:notes01}, respectively. We summarize our findings in Section \ref{sec:summary}.

\section{Sample}
\label{sec:sample}
Recently, \citet[][]{paliya_radio_morphology24} conducted an extensive study on the radio morphology of the $\gamma$-ray sources in the 4FGL-DR4 catalog that are potentially associated with AGN. This study utilized observations from the latest radio surveys, including the Low-Frequency Array Two-Metre Sky Survey (LoTSS), Very Large Array Sky Survey (VLASS), Rapid ASKAP Continuum Survey (RACS), and Faint Images of the Radio Sky at Twenty cm (FIRST) survey. They first identified objects showing a bipolar radio morphology, indicative of the misaligned nature. They also considered the overall radio spectral index, optical spectral properties, and core dominance to select the most promising \gm-ray emitting radio galaxies. These objects usually have steep radio spectra ($\alpha <-0.5$, $F_\nu \propto \nu^\alpha$), an optical spectrum dominated by host galaxy absorption features, and low core dominance \citep[$<1$; see,][for details]{paliya_radio_morphology24}. This exercise led to the identification of 149 sources. Among these, 41 are already reported in the 4FGL-DR4 as radio galaxies, while 108 have been identified as misaligned AGN for the first time. Furthermore, there are 11 misaligned AGN present in the 4FGL-DR4 catalog that are missing from the sample of \citet[][]{paliya_radio_morphology24}, likely due to them not qualifying the criteria adopted in their work. We include all of them, resulting in a final sample of 160 \gm-ray emitting radio galaxies.

\section{\textit{Fermi}-LAT Data Reduction}
\label{sec:fermi}
We analyze the \textit{Fermi}-LAT data collected for over $\sim$16 years from 4 August 2008 to 4 October 2024. A binned likelihood analysis is performed in the energy range of 100 GeV to 2 TeV using \texttt{fermipy} version 1.3.1 with \texttt{fermitools-2.2.0}. The choice of the lower energy threshold of 100 GeV is primarily driven by the fact that the VHE band is conventionally defined above 100 GeV \citep[cf.][]{2022ApJ...933..213D}. It is in this energy range of 100 GeV to tens of TeV that the current IACTs are most sensitive. Therefore, objects significantly detected above 100 GeV would be the best targets for the follow-up observations with the ground-based IACTs. We select Pass 8 \texttt{\small{SOURCE}} class events from a circular region with a radius of 2$^\circ$ around each radio galaxy in our sample. The data are spatially binned with 0.05$^\circ$ per pixel and divided into ten logarithmically spaced bins per energy decade. To ensure the data quality, we use the recommended time filter `\texttt{DATA\_QUAL > 0 \&\& LAT\_CONFIG == 1}' and apply a maximum zenith angle cut at 105$^\circ$ to avoid contamination from secondary $\gamma$-rays from the Earth's limb.
\par
For accurate modeling of the sources, the model file includes all 4FGL-DR4 sources within 5$^\circ$ of the target. The diffuse background model templates \texttt{gll\_iem\_v07} and \texttt{iso\_P8R3\_SOURCE\_V3\_v1} are used to model the Galactic diffuse and isotropic emissions, respectively. Additionally, the emissions from extended sources are accounted for using a 14-year archival catalog of extended sources\footnote{\url{https://fermi.gsfc.nasa.gov/ssc/data/access/lat/14yr_catalog/}}. Given the low photon statistics above 100 GeV, we adopt a power-law spectrum for all sources. Furthermore, we use the tool \texttt{gtsrcprob} to compute the energy of the highest-energy photon detected from the source of interest and the corresponding probability of association.
\par
At very high energies, $\gamma$-ray photons can interact with diffuse Extragalactic Background Light (EBL), leading to pair production. To account for this, we also perform an EBL-corrected spectral fitting using the EBL model proposed by \cite{ebl_saldana21} and derive the EBL-corrected spectral parameters. This was done by utilizing the EBL models available within the Fermitools, which can be inserted in the source model for likelihood fitting\footnote{\url{https://fermi.gsfc.nasa.gov/ssc/data/analysis/scitools/source\_models.html\#ExpCutoff}} \citep[e.g.,][]{2018Sci...362.1031F}.

\begin{table*}
\caption{The list of \gm-ray emitting misaligned jetted AGN detected in the VHE band. The column details are as follows: (1) 4FGL name; (2) counterpart name; (3) redshift (4), (5), and (6) observed \gm-ray photon flux (in 10$^{-12}$ \phflux), photon index and TS (in 0.1$-$2 TeV band); (7), (8), and (9) EBL-corrected \gm-ray photon flux (in 10$^{-12}$ \phflux), photon index and TS (in 0.1$-$2 TeV band); and (10) whether the source is present in TeVCat. $\dagger$: For J0308.4+0407, we also repeated the likelihood analysis by fixing the photon index to the average value found for all sources\label{tab:vhe12}}
\begin{tabular}{llcccccccc}
\hline
4FGL Name & Other name & $z$ & Flux$_{\rm obs}$ & $\Gamma_{\rm obs}$ & TS$_{\rm obs}$ & Flux$_{\rm EBL}$ & $\Gamma_{\rm EBL}$ & TS$_{\rm EBL}$ & TeV\\
~[1] & [2] & [3] & [4] & [5] & [6] & [7] & [8] & [9] & [10]\\
\hline
J0308.4+0407 &	NGC 1218 & 0.029 & $2.02\pm1.46$	 & 10.00$\pm$0.01 & 21.53 	 & $2.02\pm1.42$ &10.00$\pm$0.02 & 21.53 & N\\
     "      & "  & " & $3.28\pm2.32$ & 2.63$^\dagger$	 & 19.85 	 & - 	 & -& -& -\\
J0316.8+4120 	& IC 310 & 0.019 & $8.49\pm3.65$	 & 1.79$\pm$0.59	 & 35.76 	 & $8.48\pm3.64$ &1.71$\pm$0.60 & 35.61 & Y\\
J0319.8+4130 	& NGC 1275 & 0.018 & $22.20\pm5.68$	 & 3.06$\pm$0.57	 & 148.68 	 & $22.20\pm5.69$ 	 &3.01$\pm$0.57 	 & 148.70 & Y\\
J0334.3+3920 	& 4C +39.12 & 0.020 & $5.28\pm2.86$	 & 2.23$\pm$0.86	 & 22.33 	 & $5.28\pm2.86$ 	 &2.15$\pm$0.87 	 & 22.40 & N\\
J0550.5$-$3216 	& PKS 0548$-$322 & 0.069 & $6.46\pm3.12$	 & 2.77$\pm$0.94	 & 33.54 	 & $6.47\pm3.12$ 	 &2.54$\pm$0.98 	 & 33.74 & Y\\
J0627.0$-$3529 	& PKS 0625$-$35 & 0.055 & $11.03\pm3.96$	 & 2.63$\pm$0.68	 & 68.13 	 & $11.03\pm3.96$ 	 &2.45$\pm$0.70 	 & 67.77 & Y\\
J0912.9$-$2102 	& MRC 0910$-$208 & 0.198 & $15.90\pm5.07$	 & 2.64$\pm$0.60	 & 77.06 	 & $15.90\pm5.09$ 	 &1.89$\pm$0.68 	 & 78.12 &  Y\\
J1230.8+1223 	& M 87 & 0.004 & $12.91\pm4.72$	 & 2.38$\pm$0.62	 & 66.28 	 & $12.89\pm4.72$ 	 &2.38$\pm$0.62 	 & 66.28 & Y\\
J1310.6+2449 	& GB6 B1308+2504 & 0.226 & $4.23\pm2.46$	 & 3.38$\pm$1.47	 & 27.91 	 & $4.27\pm2.48$ 	 &2.72$\pm$1.64 	 & 28.10 & N\\
J1325.5$-$4300 	& Centaurus A & 0.002 & $11.60\pm4.28$	 & 2.76$\pm$0.73	 & 59.65 	 & $11.60\pm4.28$ 	 &2.76$\pm$0.73 	 & 59.65 & Y\\
J1341.2+3958 	& GB6 B1338+4014 & 0.172& $7.64\pm3.27$	 & 2.51$\pm$0.75	 & 45.98 	 & $7.66\pm3.28$ 	 &1.85$\pm$0.84 	 & 46.48 & N\\
J1449.5+2746 	& B2 1447+27 & 0.031 & $4.04\pm2.36$	 & 2.40$\pm$1.01	 & 24.07 	 & $4.04\pm2.37$ 	 &2.29$\pm$1.03 	 & 24.14 & N\\
\hline
\end{tabular}
\end{table*}

\section{Results and Discussion}
\label{sec:results}
The \textit{Fermi}-LAT data analysis of a sample of 160 radio galaxies covering the energy range 0.1$-$2 TeV, results in the detection of 12 sources with TS$>$16, of which nine have TS$>$25 (Table \ref{tab:vhe12}). Among these 12 objects, seven were previously identified as VHE-emitting radio galaxies using ground-based Cherenkov telescopes, and all but one have been classified as FR-I-type sources. The only known FR-II with VHE emission, 4FGL J0912.9$-$2102, was originally reported to be a BL Lac object; however, recently, it was reclassified as an FR-II radio galaxy (Section \ref{sec:known}). We report two new VHE-emitting radio galaxies, and three additional promising candidates with 16$<$TS$<$25.

In the sample of 12 VHE-emitting radio galaxies, only two objects exhibit FR-II-shaped radio morphology, and the remaining ten sources show diffuse, low-surface brightness extended radio emission typically seen in FR-I radio sources. The cause of larger number of VHE-emitting FR-I sources in unclear. It could be simply due to FR-Is being located closer with respect to FR-II objects \citep[see,][]{2018NatSR...815097G,2018MNRAS.475.3429W,paliya_radio_morphology24}. Another possibility could be that FR-I sources may live in denser environments whose interaction with the jet may provide necessary conditions for accelerating particles to VHE \citep[e.g.,][]{2018ApJ...864..118K,2021ApJ...920L..24K,2022Galax..10...61R}. 

We also identify 13 additional radio galaxies with at least one $>$100 GeV photon detected with source probability greater than 95$\%$ (Table \ref{tab:hep}). This subset includes five FR-II, five FR-I, and three radio galaxies with hybrid morphology. These potential VHE emitters present promising targets for ground-based IACTs and the upcoming CTAO, as none of them are present in TeVCat. Furthermore, two or more photons are detected from all 12 VHE-detected radio galaxies, including 14 VHE photons from NGC~1275 (Table \ref{tab:hep}). A 203 GeV photon was detected from 4FGL J1258.7$+$5143, an FR-II radio galaxy located at the redshift of 0.464, suggesting it to be the farthest VHE-emitting radio galaxy. Among the sources detected with TS$>$16, 4FGL J1310.6$+$2449 ($z=0.226$) is the most distant FR-II radio source.

\begin{figure*}
    \hbox{
    \includegraphics[scale=0.52]{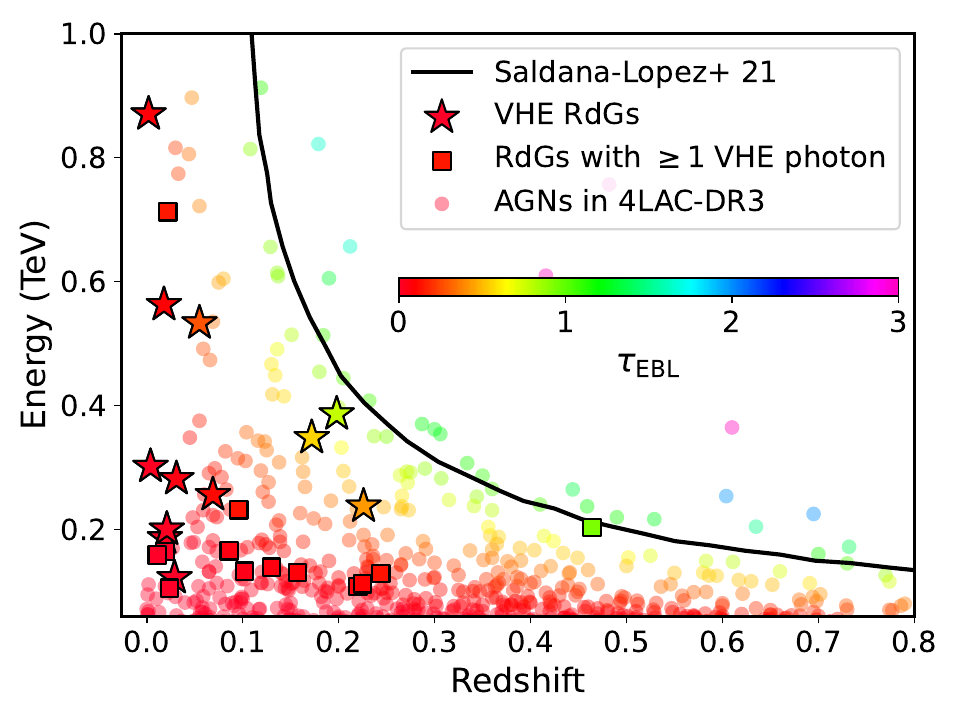}
    \includegraphics[scale=0.48]{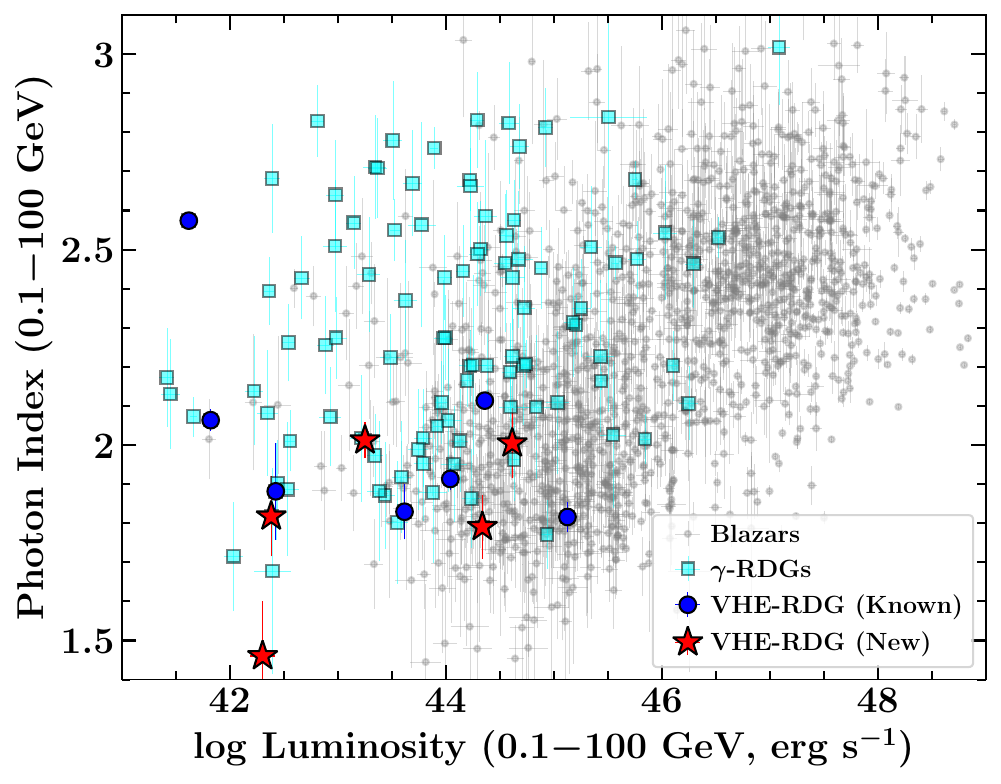}
    }
    \caption{Left: The energy of the highest-energy photon versus redshift for each source in Table \ref{tab:hep} is plotted. The stars represent the radio galaxies with TS $>$ 16, and squares represent objects with at least one VHE photon detected from them. The cosmic $\gamma$-ray horizon is shown as predicted by the EBL model proposed by \citet[][]{ebl_saldana21}. Right: This plot shows the variation of the \gm-ray photon index with \gm-ray luminosity.}
    \label{fig:cosmic_horizon}
\end{figure*}

EBL correction is crucial for extragalactic $\gamma$-ray sources emitting at energies above 100 GeV, particularly at redshifts above 0.1. As these VHE photons travel through the Universe, they interact with the EBL, leading to energy-dependent attenuation due to pair production. Therefore, EBL attenuation is taken into account to derive the intrinsic spectral parameters for the sources. The observed and EBL-corrected parameters for each source are listed in Table \ref{tab:vhe12}. The intrinsic \gm-ray spectral shapes appear harder, especially for targets at higher redshifts. The left panel of Figure \ref{fig:cosmic_horizon} shows the highest-energy photon plotted against redshift for the sources listed in Table \ref{tab:hep}, in agreement with the predicted Cosmic Gamma-Ray Horizon (CGRH) based on the latest EBL model by \cite{ebl_saldana21}. The CGRH defines the energy-redshift relationship where the Universe becomes opaque to $\gamma$-rays, corresponding to an optical depth of $\tau_{\rm EBL} (E,z) = 1$.

\begin{table}
\caption{List of radio galaxies with at least one photon detected above 100 GeV with $\geq$ 95 \% source probability. The 12 VHE emitters from Table \ref{tab:vhe12} are listed in the top panel. The column details are as follows: (1) 4FGL name; (2) redshift; (3) number of detected VHE photons; (4) energy of the highest energy photon (in GeV); (5) time of VHE photon detection (in MJD); and (6) morphological classification \citep[adopted from][]{paliya_radio_morphology24,2025arXiv250703105P}.\label{tab:hep}}
\begin{tabular}{lccccc}
\hline
4FGL Name & $z$ & No. & $E_{\rm HEP}$ & $T_{\rm arrival}$ & Morph.\\
~[1] & [2] & [3] & [4] & [5] & [6]\\
\hline
 J0308.4+0407 & 0.029& 2 & 121.90 & 58400.33 & FR I\\
 J0316.8+4120 & 0.019& 3 & 186.28 & 60031.22 & FR I\\
 J0319.8+4130 & 0.018& 14 & 562.73 & 57983.04 & FR I\\
 J0334.3+3920 & 0.020& 2 & 200.17 & 58381.70 & FR I\\
 J0550.5$-$3216 & 0.069& 4 & 255.81 & 59609.96 & FR I\\
 J0627.0$-$3529 & 0.055& 7 & 533.13 & 55520.59 & FR I\\
 J0912.9$-$2102 & 0.198& 8 & 386.16 & 58545.01 & FR II\\
 J1230.8+1223 & 0.004& 6 & 301.48 & 57234.41 & FR I\\
 J1310.6+2449 & 0.226& 3 & 237.43 & 55856.06 & FR II\\
 J1325.5$-$4300 & 0.002& 5 & 870.40 & 59392.08 & FR I\\
 J1341.2+3958 & 0.172& 5 & 348.07 & 60261.13 & FR I\\
 J1449.5+2746 & 0.031& 3 & 281.67 & 58970.07 & FR I\\
\hline
J0153.4+7114 & 0.022& 2 & 712.68 & 57005.52 & FR I\\
J0627.0+2623 & 0.157& 1 & 129.97 & 54721.75 & FR I-II\\
J0708.9+4839 & 0.019& 1 & 163.87 & 58672.94 & FR I\\
J1149.0+5924 & 0.011& 1 & 158.38 & 60115.68 & FR I\\
J1258.7+5143 & 0.464& 1 & 202.89 & 55459.62 & FR II\\
J1306.3+1113 & 0.086& 1 & 165.41 & 57040.74 & FR I\\
J1402.6+1600 & 0.244& 1 & 128.72 & 56971.14 & FR II\\
J1512.2+0202 & 0.220& 1 & 107.72 & 60181.78 & FR II\\
J1516.8+2918 & 0.130& 1 & 138.87 & 56279.69 & FR II\\
J1518.6+0614 & 0.102& 1 & 132.47 & 56615.09 & FR I\\
J1630.6+8234 & 0.024& 1 & 105.03 & 60153.84 & FR I-II\\
J1644.2+4546 & 0.225& 1 & 112.10 & 57455.94 & FR II\\
J2359.3$-$2049 & 0.096& 1 & 231.56 & 59083.33 & FR I-II\\
\end{tabular}
\end{table}

In the right panel of Figure~\ref{fig:cosmic_horizon}, we show the variation of the \gm-ray photon index as a function of the \gm-ray luminosity for Fermi-LAT detected AGN. In this diagram, newly VHE-detected radio galaxies tend to occupy a region of hard \gm-ray spectra (photon index $<$ 2) and low \gm-ray luminosity ($<1\times10^{45}$ \lum), similar to other misaligned AGN detected previously with IACTs. This observation indicates the \gm-ray production mechanism in new VHE emitters is similar to that at work in the known VHE-detected radio galaxies.

\begin{figure*}
    \centering
    \hbox{
    \includegraphics[scale=0.2]{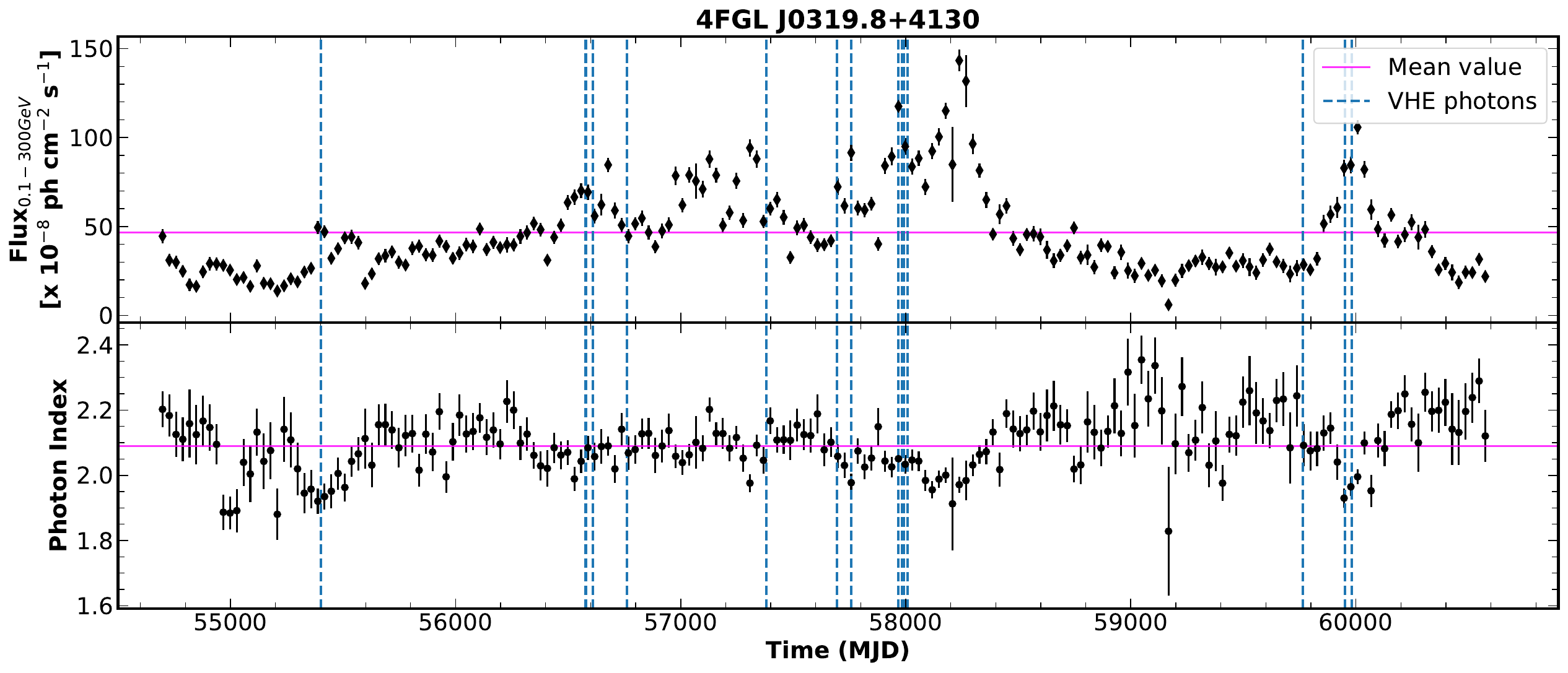}
    \includegraphics[scale=0.2]{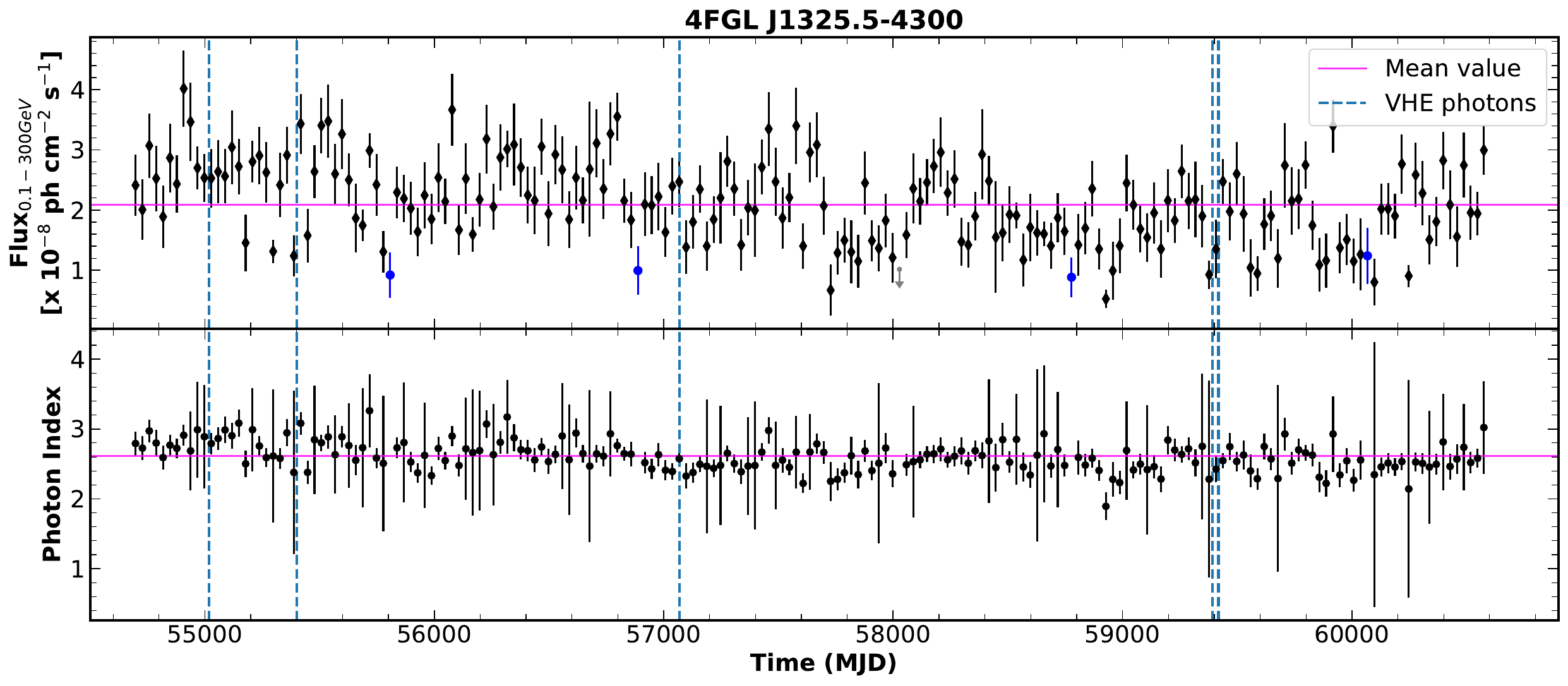}
    }
    \caption{Lightcurves of two VHE sources, in the 100 MeV$-$300 GeV energy band constructed using \textit{Fermi}-LAT data. The blue dashed vertical lines represent the time of arrival of VHE photons. For the bins where the photon index could not be estimated, the blue circles represent the flux computed by fixing the photon index at the average value. The light curves of the remaining objects are provided in the Appendix (Figure~\ref{fig:app}).}
    \label{fig:lc}
\end{figure*}

The VHE observations of jetted AGN by IACTs are usually triggered by flaring activity detected at lower energies \citep[cf.][]{2022ApJ...924...95A}. In fact, the first VHE detections of several radio galaxies were reported at the time of elevated activity episodes \citep[e.g.,][]{ic310_magic_apj_2010,aleksic_magic_ngc1275_2012}. Therefore, we examine the temporal \gm-ray flux variations of all 12 sources with TS$>$16 to ascertain whether Fermi-LAT also detected VHE photons during epochs of flaring activity. We generate monthly-binned \gm-ray light curves in the energy range of 0.1$-$300 GeV following the standard data reduction procedure considering 15$^{\circ}$ region of interest and adopting all \gm-ray sources lying within 20$^{\circ}$ from the position of the target source. We keep the spectral parameters of all bright sources (TS$>$9) free to vary during the fit. {In the time bins with fit non-convergence}, we freeze the photon index to its mission-averaged value and allow only the normalization parameter to vary. For NGC 1275 and Cen A, we have generated monthly-binned light curves and show their flux and photon index variations in Figure~\ref{fig:lc}. For all other sources, three-month binning is adopted due to low photon statistics (see Figure~\ref{fig:app}). In all the light curves, we overplot the arrival time of the VHE photons with vertical dashed lines. 

For a majority of sources, we do not find any elevated activity periods at the time of the VHE photon detection. Interestingly, NGC 1275  exhibits a harder-when-brighter trend and several VHE photons are detected during its elevated activity episodes. This result is consistent with those reported for other Fermi-LAT detected AGN \citep[e.g.,][]{2011ApJ...730L...8A}. The details about individual objects are provided in the next section and the Appendix.

\section{Notes on Individual Objects}
\label{sec:notes01}
\subsection{Newly identified VHE-emitting radio galaxies}
\textit{4FGL J0308.4+0407 (NGC 1218)}: 
This object is classified as an FR-I radio galaxy \citep{1986MNRAS.219..545S,balmaverde21} hosting a supermassive black hole (SMBH) of mass 5.45 $\times$ 10$^8$ M$_\odot$ and is located at a distance of 116 Mpc \citep{falcke04}. This source is known to exhibit an extended optical and X-ray jet \citep[][]{1995ApJ...450L..55S,2015ApJS..220....5M}. No significant flux variability is seen in the \textit{Fermi}-LAT lightcurve, and two VHE photons are detected when the \gm-ray flux level is similar to the mission-averaged value (Figure \ref{fig:app}). Due to poor photon statistics, the photon index of the fitted power law model could not be estimated. Therefore, we repeat the likelihood fitting by freezing it to 2.63, i.e., the average photon index for all other detected sources. In both cases, the estimated TS is $>$16 (Table~\ref{tab:vhe12}), indicating that NGC 1218 is a promising VHE emitting candidate. This object is not present in the TeVCat and was proposed as a promising VHE-emitting candidate for CTAO detection by \citet[][]{rutlen20}. 

\textit{4FGL J0334.3+3920 (4C +39.12)}: 
It is classified as an FR-I radio galaxy \citep{1986A&AS...64..135P,4c_jet_2001} and is a faint \gm-ray emitting source. 
It harbors an SMBH of mass 1.61 $\times$ 10$^8$ M$_\odot$ and is located at a distance of 85.8 Mpc \citep{4c_ks}. It shows a parsec-scale jet with possible limb-brightened structure \citep{4c_jet_2001}. In the \textit{Fermi} energy band, it shows significant $\gamma$-ray excess above 30 GeV, and the spectral energy distribution characteristics are found to be similar to that of 3C~264 using 10 years of \textit{Fermi} data \citep{rutlen20}. The \textit{Fermi}-LAT detects two VHE photons during periods of low activities (Figure \ref{fig:app}).

\textit{4FGL J1310.6+2449 (CRATES J131038.52+244822.1)}:
This object is classified as a BL Lac in the 4FGL-DR4 catalog. However, \citet{paliya_radio_morphology24} identify an FR-II morphology of the source based on the FIRST and LOFAR observations. Additionally, its optical spectrum shows host galaxy absorption features, thus supporting the radio galaxy nature \citep{2020Ap&SS.365...12D}. Interestingly, it has been proposed as a potential neutrino source due to its close proximity with the IceCube event IC200921A \citep{atel_crates_garrappa2020}. The \textit{Fermi}-LAT detects 3 VHE photons well before the IceCube neutrino event in 2011, 2017, and 2018. During the arrival time of all three VHE photons, the source flux level remains comparable to its mission-averaged value. It is the most distant radio galaxy detected at VHE energies in our sample and one of the only two VHE-emitting FR-II sources.  

\textit{4FGL J1341.2+3958 (SDSS J134105.10+395945.4)}:
It was formerly classified as a high-synchrotron peaked blazar \citep{4fgl_dr4, hbl24}. However, it was later identified as a low-power radio galaxy by \cite{lin2018}. Recent work by \cite{paliya_radio_morphology24} further confirmed its radio galaxy morphology to resemble an FR-I source, displaying wide-angle tail features visible in the LOFAR image. The source hosts an SMBH of mass $\sim 8 \times 10^{8} M_\odot$ \citep{hbl24}. This VHE-emitting radio galaxy is a faint \gm-ray emitter and was barely detected only on a few occasions during the 16 years of the \textit{Fermi}-LAT operation. No elevated \gm-ray flux activity was seen at the time of the arrival of VHE photons (Figure \ref{fig:app}). The \textit{Fermi}-LAT detection of VHE photons from this source aligns with the prediction by \cite{te_rex}, who suggested its visibility in relatively shallow observation by CTAO or 50-hour observation by MAGIC. 

\textit{4FGL J1449.5+2746 (B2 1447+27)}:
This source is an FR-I radio galaxy that is located at a redshift of 0.031 \citep{4fgl_dr4, foschini22}. It hosts an SMBH of mass $\sim 7.2 \times 10^8 M_\odot$.  Interestingly, this radio galaxy is found to be in directional correlation with a bronze track-like neutrino event IC141114A \citep{icecat23}. The three VHE photons from this source arrived during period of average activity as revealed by the three-month binned \textit{Fermi}-LAT lightcurve (Figure \ref{fig:app}).

\subsection{Known VHE source missing in our sample}
4FGL J1144.9$+$1937 or 3C 264 is an FR-I type radio galaxy located at a redshift of 0.021
\citep[][]{1981AJ.....86.1165B,2004A&A...415..905L}.
3C 264 is the only previously known VHE-emitting radio galaxy that is missing from the list of VHE emitters presented in this work. The highest photon energy recorded from 3C 264 by the \textit{Fermi}-LAT is $\sim$97 GeV \citep[][]{4lac_22}, which explains the non-detection in the LAT data analysis. It was significantly detected ($>$5$\sigma$ c. l.) with VERITAS during observing run from 2018 March 9-21 (total exposure 17.7 hours) when it was undergoing a flaring activity \citep[][]{2020ApJ...896...41A}. On 2018 March 16, Fermi-LAT faced the Y $-$ solar array drive assembly anomaly and resumed taking observations only on 2018 April 8\footnote{\url{https://www.nasa.gov/centers-and-facilities/goddard/fermi-status-update/}}. This implies that Fermi-LAT could not fully cover the VHE elevated activity episode from 3C 264.

\section{Summary}
\label{sec:summary}

In this study, we utilize over 16 years of \textit{Fermi}-LAT data to explore VHE emission from the recently updated catalog of \gm-ray emitting radio galaxies. While VHE sky is traditionally studied by ground-based Cherenkov telescopes, e.g., MAGIC, VERITAS, and H.E.S.S., their limitations in sky coverage and observing conditions leave this energy range relatively under-explored. \textit{Fermi}-LAT, operating in all-sky mode for over 16 years, offers complementary capabilities up to $\sim$ 2 TeV energies, making it a powerful tool for identifying bright VHE sources.

Out of 160 sources, we identify nine misaligned jetted objects significantly detected in the VHE band at high confidence (TS $>$ 25), and three additional promising candidates (16$<$TS$<$25). Out of eight previously known VHE-emitting radio galaxies, seven of them are present in our sample, thus indicating the effectiveness of the \textit{Fermi}-LAT in discovering the VHE emission from sources of this class. We also report, for the first time, the detection of VHE emission from two radio galaxies and three promising candidates, thus significantly expanding the small sample of such objects. Additionally, 13 radio galaxies are identified as potential VHE emitters, marking them as promising candidates for future observations by upcoming CTAO. 

We generate \gm-ray light curves in 0.1$-$300 GeV energy range to examine any possible trend in the temporal/spectral flux variability with the arrival time of the VHE photons. We find that 4FGL J0319.8+4130 or NGC 1275 exhibits flaring activity during the detection of the VHE photons, and a `harder-when-brighter' trend. However, for other sources, no significant brightening is noticed at the time of the VHE photon arrival. 

The discovery of VHE radiation from misaligned jetted AGN population, despite weak Doppler boosting, suggests complex emission processes and different emission sites than known for blazars. Identifying more radio galaxies in the VHE sky and subsequent multiwavelength follow-up observations will help us improve our understanding of VHE $\gamma$-ray sky.

\section*{Acknowledgements}
We thank the journal referee for constructive criticism and comments which have helped improve the manuscript. 
This research has made use of NASA's Astrophysics Data System Bibliographic Services. The use of the \textit{Fermi}-LAT data provided by the Fermi Science Support Center is gratefully acknowledged.

\section*{Data Availability}
The data utilized in this study are publicly accessible and were obtained from the archives available at \url{https://Fermi.gsfc.nasa.gov/}.

\newpage

\appendix
\section{Gamma-ray light curves}
\begin{figure*}
    \centering
    \hbox{
    \includegraphics[scale=0.21]{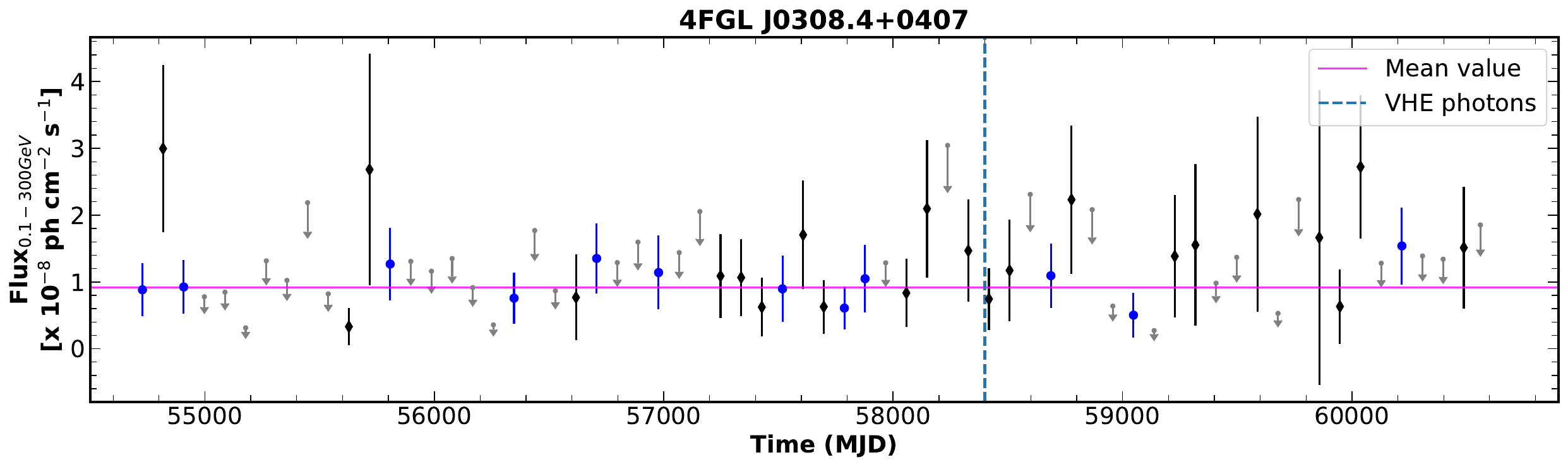}
    \includegraphics[scale=0.21]{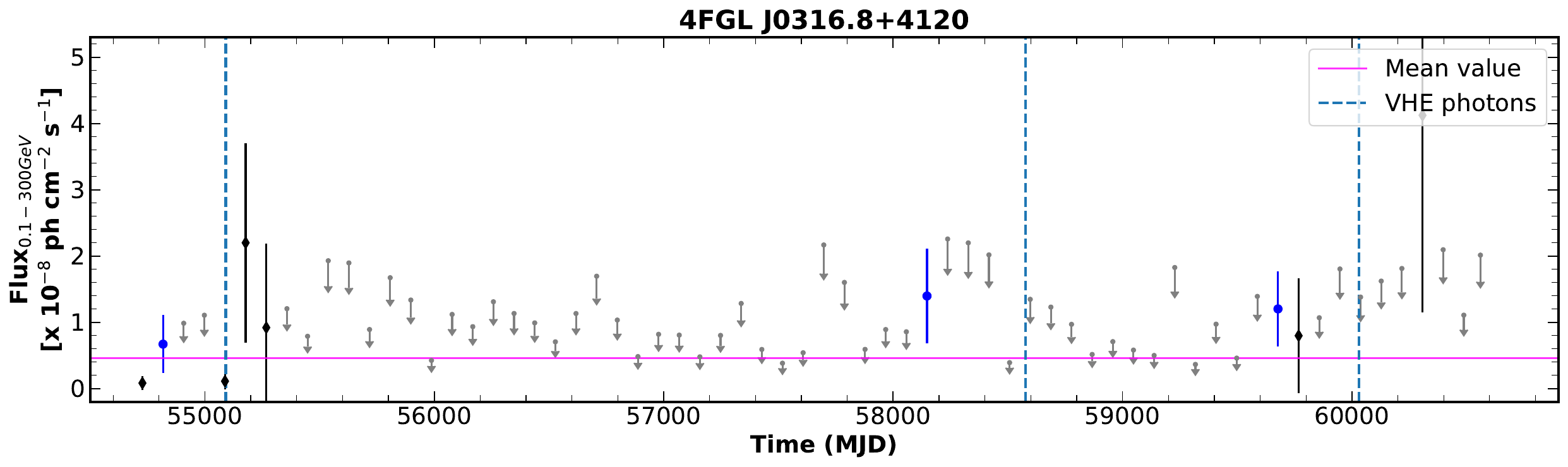}
    }
    \hbox{
    \includegraphics[scale=0.21]{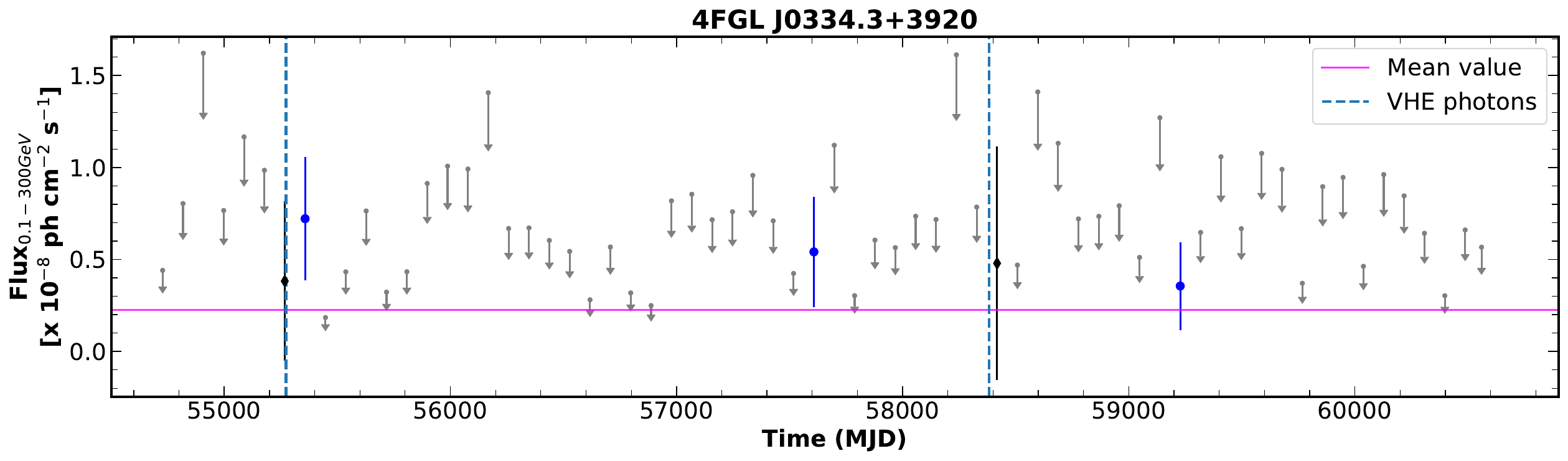}
    \includegraphics[scale=0.21]{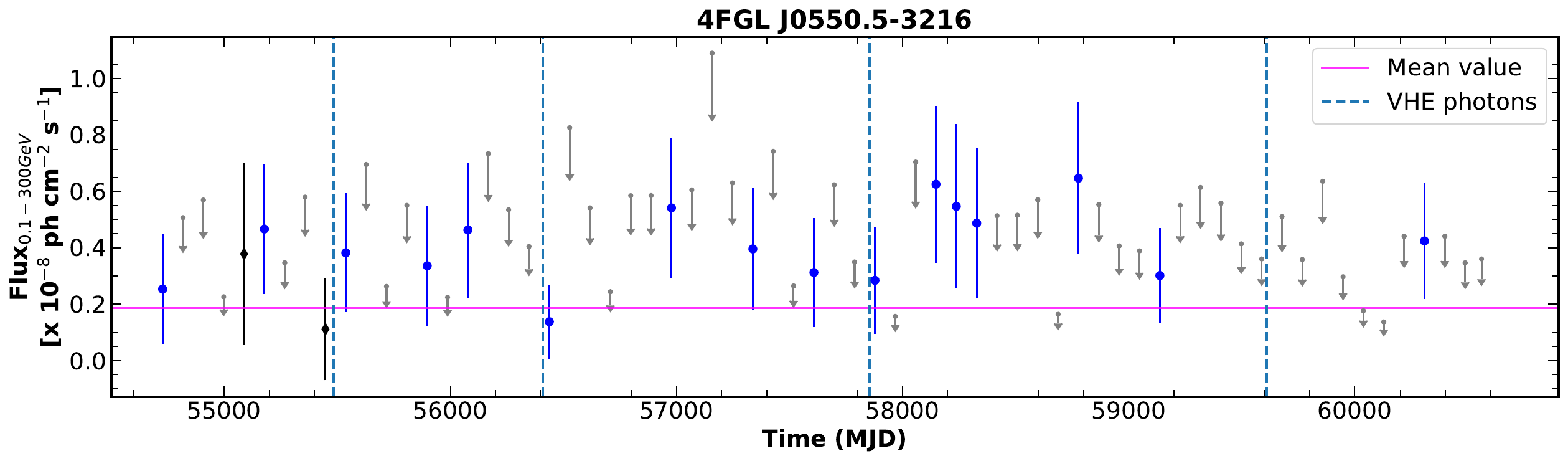}
    }
    \hbox{
    \includegraphics[scale=0.21]{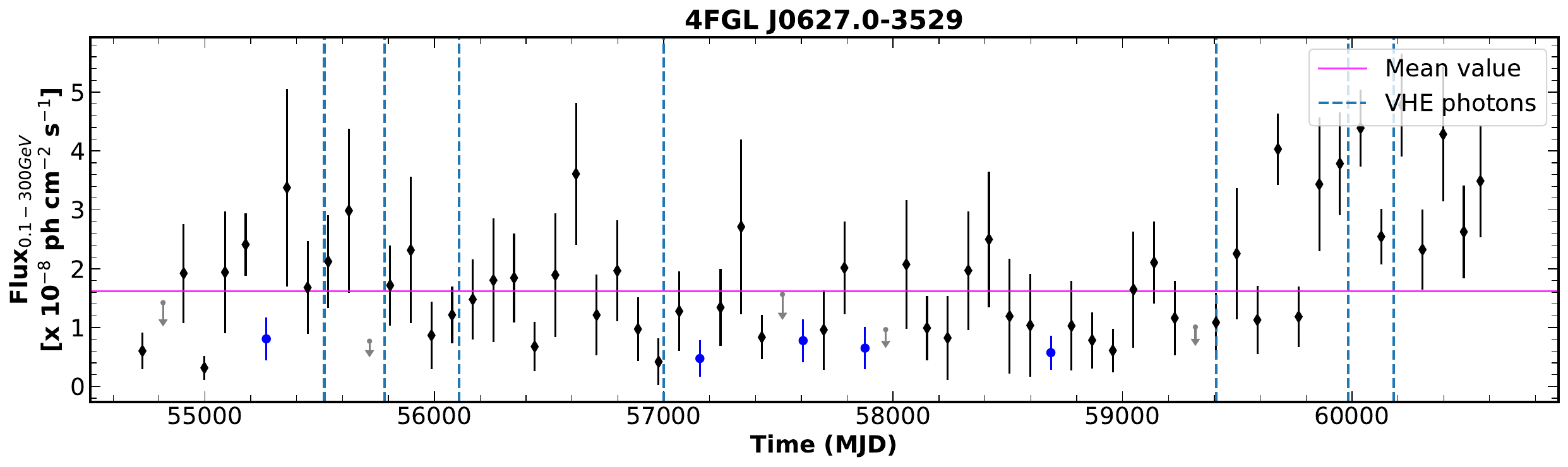}
    \includegraphics[scale=0.21]{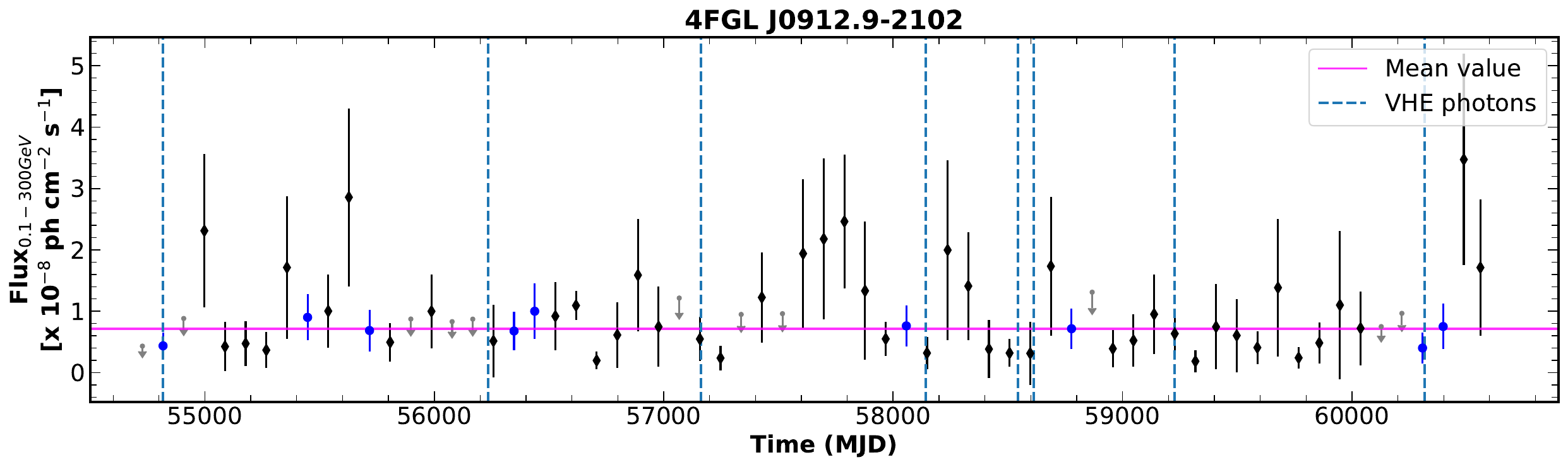}
    }
    \hbox{
    \includegraphics[scale=0.21]{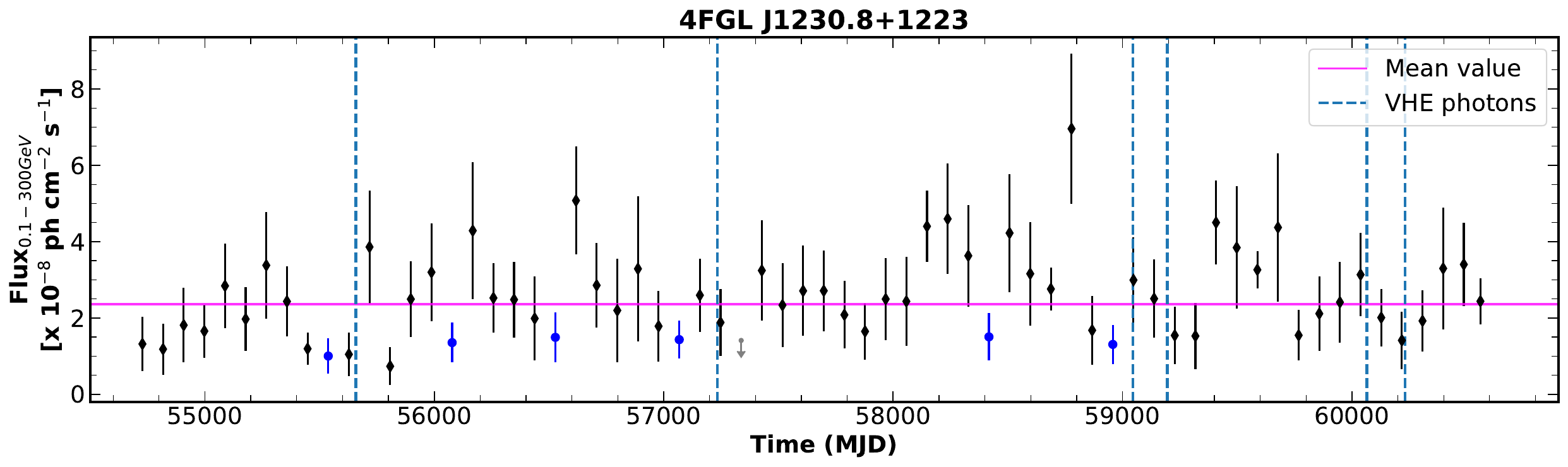}
    \includegraphics[scale=0.21]{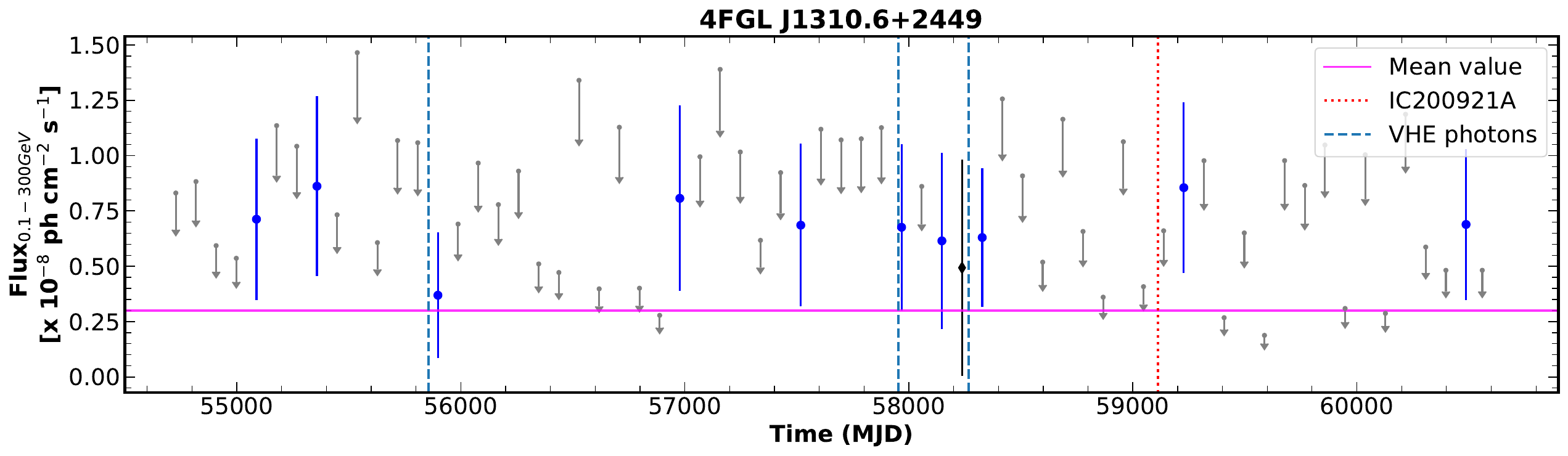}
    }
    \hbox{
    \includegraphics[scale=0.21]{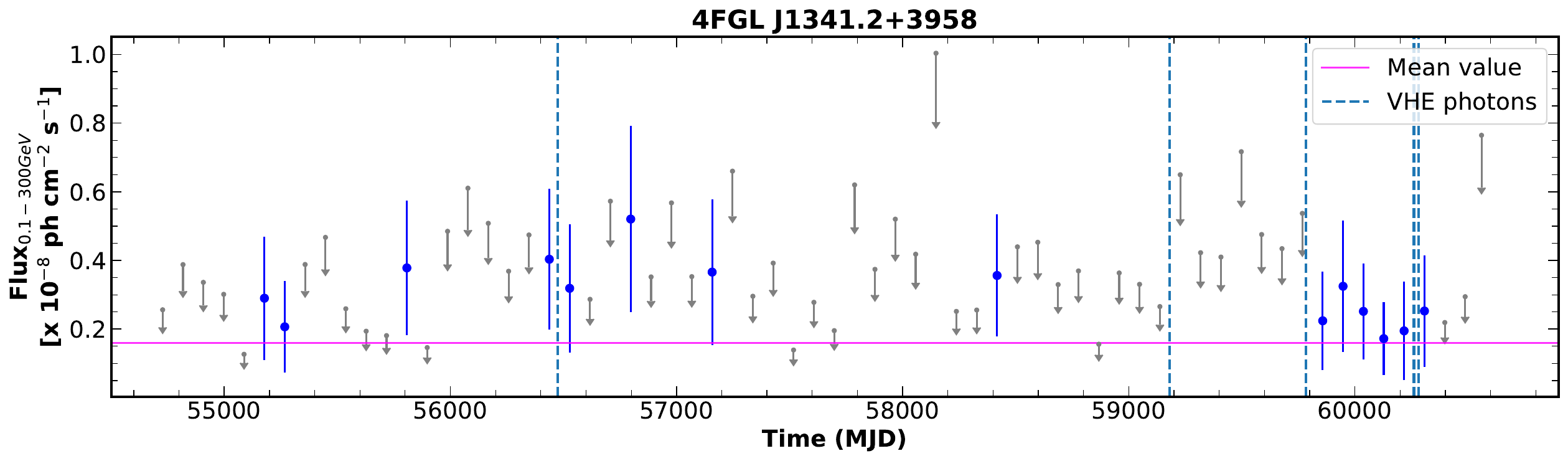}
    \includegraphics[scale=0.21]{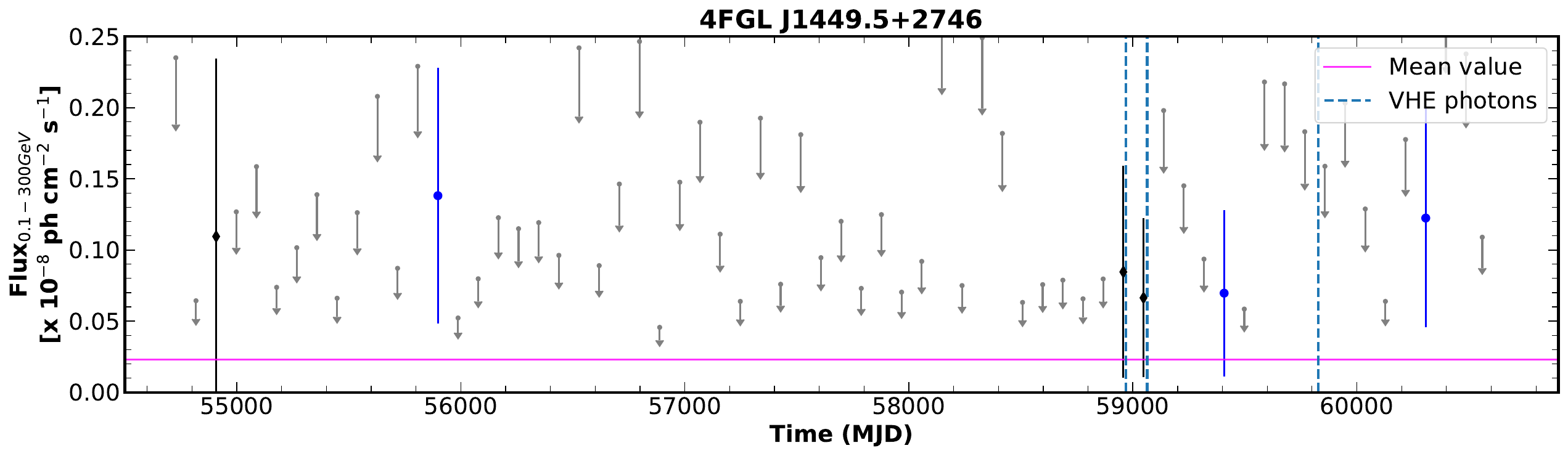}
    }
    \caption{Lightcurves of 10 VHE sources. Other information are same as in Figure~\ref{fig:lc}.}
    \label{fig:app}
\end{figure*}

The three-month binned \gm-ray light curves of 10 misaligned AGN are shown in Figure~\ref{fig:app}.

\section{Known VHE-emitting radio galaxies}
\label{sec:known}
\textit{4FGL J0316.8+4120 (IC 310)}: 
It is an FR-I radio galaxy located in the Perseus cluster \citep[][]{2020MNRAS.499.5791G}. It is a known VHE emitter, first discovered by MAGIC during 2009-2010 while observing NGC~1275, which is located 0.6$^\circ$ away \citep{magic2010,ic310_magic_apj_2010}. The GeV$-$TeV emission was observed to be variable on timescales as short as 5 minutes \citep{ic310_variable_aleksic_2014}, similar to blazars. A harder-when-brighter behavior was also seen in X-rays and VHE emission akin to blazars \citep{ahnen2017}. Recently, \citet[][]{2020MNRAS.499.5791G} observed this object using the Very Large Array and proposed the possible jet bending with the inner jet lying close to the line of sight to the observer. Furthermore, VERITAS and Large High Altitude Air Shower Observatory (LHAASO) have also reported variable VHE emission from this source \citep{lhaaso24,ic310_veritas_24}.

\textit{4FGL J0319.8+4130 (NGC 1275)}:
This object is a \gm-ray bright FR-I radio source at the center of the Perseus cluster. The radio structure exhibits a prominent core with diffuse outer lobes \citep[][]{1990MNRAS.246..477P,2020MNRAS.499.5791G}.
VHE emission from this source was discovered in July 2010 by MAGIC during an increased $\gamma$-ray activity reported by \textit{Fermi}-LAT \citep{aleksic_magic_ngc1275_2012}. This object has also been detected by VERITAS and recently by the Major Atmospheric Cherenkov Experiment Telescope (MACE) and LHAASO \citep{veritas_ngc1275,godambe24_ngc1275, lhaaso_ngc1275}. Variability across different timescales has been observed in \gm~rays, while MAGIC reported significant variability in the day-to-day VHE lightcurve, with flux doubling time of $\sim$0.4 day \citep{ngc1275_magic18}. A single-zone SSC model has been successful in explaining the multiwavelength SED with a modest Doppler factor in the range of 2.0-4.3 \citep{aleksic_magic_ngc1275_2012,fukazawa18, ngc1275_magic18, gulati21, godambe24_ngc1275}. The \textit{Fermi}-LAT has detected 14 VHE photons from this object. The arrival times of a majority of them have been found to be coincident with the elevated \gm-ray flux activity periods. The \gm-ray spectrum during these epochs is flatter than that noticed from 16-year averaged data analysis.

\textit{4FGL J0550.5$-$3216 (PKS 0548$-$322)}: 
It was classified as a BL Lac in the 4FGL-DR4 catalog. However, it exhibits a wide-angled tailed FR-I morphology in both RACS and VLASS images and low core dominance. The optical spectrum shows host galaxy absorption features \citep{fosbury}. These observations resulted in the reclassification of this source as an FR-I radio galaxy \citep{paliya_radio_morphology24}. VHE emission from this source was detected using H.E.S.S. observations between 2004 and 2006, where the SSC component was found to explain the HE-VHE bump well with a Doppler factor of 10, higher than the values corresponding to the radio galaxies \citep{pks0548_vhe_22}. It is a faint \gm-ray emitter and 
has been significantly detected only on a few occasions in the last $\sim$16 years of the \textit{Fermi}-LAT operation. 

\textit{4FGL J0627.0$-$3529 (PKS 0625$-$35)}:
This object is a known VHE emitter \citep{dyrda625, hess_625}. It shows an FR-I morphology with kiloparsec jets and diffuse extended lobes \citep{ojha10, paliya_radio_morphology24}. This object was classified as a radio galaxy in the 4FGL-DR4 catalog. Variability in both the MeV$-$GeV and GeV$-$TeV bands were reported, including a VHE  outburst in November 2018 \citep{pks0625_fermi2015,pks0625_hess_iagn_2024}. The \textit{Fermi}-LAT light curve reveals the source to be in its average activity state during the arrival times of VHE photons detected with the \textit{Fermi}-LAT.

\textit{4FGL J0912.9$-$2102 (MRC 0910$-$208)}:
It has previously been classified as a high-synchrotron peaked blazar \citep[HSP;][]{chang19}. However, its RACS image shows FR-II morphology with radio lobes and hotspots on opposite sides. Additionally, the optical spectrum is dominated by absorption features of the host galaxy, and a clear Ca H$\&$K break also confirms the presence of misaligned jets \citep{paliya_radio_morphology24}.
The VHE emission from this source was detected by H.E.S.S. in 2018 with a 17.2 hrs observation \citep{mrc_hess22}. With its revised classification, it is the only FR-II radio galaxy known to be a VHE emitter. Eight VHE photons have been detected with \textit{Fermi}-LAT, and no flaring activity has been identified in the light curve at the time of their detection. 

\textit{4FGL 1230.8+1223 (M87)}:
This object is an FR-I radio galaxy \citep[][]{2000ApJ...543..611O}. The VHE emission from this source was first detected in 2003 by High Energy Gamma Ray Astronomy (HEGRA) IACT \citep{m87_hegra}, which was later confirmed by H.E.S.S. \citep{m87_aharonian_2006}. There were subsequent VHE detections by MAGIC, VERITAS, and High Altitude Water Cherenkov (HAWC) Observatory \citep{m87_magic, m87_veritas, m87_hawc}. It is known to be a variable VHE $\gamma$-ray source with a  few major flares on a timescale of $\sim$1 day with possible origin from core or HST-1 knot \citep{harris}. Most recently, LHAASO monitored M87 continuously from 2021 to 2024, reporting a major VHE flare in 2022 \citep{m87_lhaaso24}. Interestingly, this flare coincided with a possible GeV flare detected by \textit{Fermi}-LAT. Several models have been proposed to explain the VHE emission observed in M87, including upstream Compton scattering in decelerating relativistic jet, spine-sheath layers, magnetospheric gaps, mini-jet model, magnetic reconnections, and hadronic models \citep{deceleration05, spinesheath_m87, m87_gap08, m87_minijet, reconnection_m87, barkov}. We do not find any episodes of the \gm-ray flaring activity at the time of the VHE detection with the \textit{Fermi}-LAT.

\textit{4FGL J1325.5$-$4300 (Centaurus~A)}:
It is one of the closest FR-I radio galaxies at a distance of $\sim 3.8$ Mpc, characterized by its highly collimated, asymmetrically edge-brightened jet and a faint counter-jet plus two giant radio lobes \citep[][and references therein]{2011ApJ...740...17F}. Gamma rays from its core were first detected in the MeV-GeV range by the Compton Gamma-Ray Observatory \citep[CGRO;][]{cgro99} and its later detection at energies above hundreds of GeV by the H.E.S.S \citep{cenA_hess} established it as a VHE emitter. Owing to its proximity, VHE emission from its kiloparsec jet has also been resolved \citep{cena_hess_2020}, thus providing clues for alternative VHE emission sites. It is also the first extended source in the GeV sky imaged by \textit{Fermi} \citep{fermi_cena_ext_2010}. Additionally, the GeV $\gamma$-ray spectrum exhibits a significant spectral hardening above $\sim$ 4 GeV, as seen by \textit{Fermi}-LAT \citep{brown2017}. Single-zone SSC models fail to explain its broadband emission \citep{cena_hess_fermi18}, leading to alternative scenarios including spine-sheath structured jet models, photo-meson decay, and extended emission processes such as proton-proton interactions \citep{sahakyan13} and inverse-Compton scattering on kiloparsec scales \citep{hardcastle11}. This object has been persistently detected in the 0.1$-$300 GeV energy range throughout the $\sim$16 years of the \textit{Fermi}-LAT operation, and all VHE photons have arrived during periods of average/low activity of the source.

\bibliographystyle{elsarticle-harv}

\end{document}